\documentclass[
    ,final            
  ]
  {aipproc}

\layoutstyle{6x9}

\begin{document}

\title{Perfect fluidity of sQGP core and dissipative hadronic corona}


\author{Tetsufumi Hirano}{
  address={Department of Physics,
   Columbia University, New York, NY 10027, USA}
}

\begin{abstract}
We can establish a new picture, 
the perfect fluid sQGP core and
the dissipative hadronic corona,
of the space-time evolution of
produced matter in relativistic heavy ion collisions
at RHIC.
It is also shown that the picture works well
also in the forward rapidity region through an analysis
based on a new class of the hydro-kinetic model
and is a manifestation of deconfinement.
\end{abstract}

\maketitle


The agreement of ideal hydrodynamic predictions
\cite{Huovinen:2003fa}
of integrated
and differential elliptic flow and radial flow patterns
with Au+Au data at RHIC energies
\cite{Ackermann:2000tr,Adcox:2002ms,Back:2002gz,Ito} 
is one of the main lines of the announcement,
``RHIC serves the perfect liquid'' \cite{BNL}.
We first study the sensitivity of this conclusion
to different hydrodynamic assumptions 
in the hadron phase.
It is found that an assumption of chemical equilibrium
neglecting viscosity in the hadron phase
in hydrodynamic simulations causes accidental reproduction
of transverse momentum spectra and 
differential elliptic flow data.
From a systematic comparison of hydrodynamic
results with the experimental data, dissipative effects
are found to be mandatory in the hadron phase.
Therefore, what is discovered at RHIC is not only
the perfect fluidity of the sQGP core
but also its dissipative hadronic corona.
Along the lines of these studies,
we develop a hybrid dynamical model
in which a \textit{fully three-dimensional}
hydrodynamic description of the QGP phase
is followed by a kinetic description of the hadron phase.
We show rapidity dependence of elliptic flow from this hybrid model
supports the above picture.
Finally, we argue that this picture is a manifestation of
deconfinement transition.

A perfect QGP fluid is assumed 
in most hydrodynamic simulations.
While one can find various assumptions in the hadron phase,
e.g.
(1) ideal and chemical equilibrium (CE) fluid,
(2) ideal and chemically frozen fluid 
(or partial chemical equilibrium, PCE), or
(3) non-equilibrium resonance gas through hadronic cascade models (HC).
Hydrodynamic results are compared with the current
differential elliptic flow data, $v_2(p_T)$,
in Fig.~20 in Ref.~\cite{Adcox:2004mh}
with putting an emphasis on the difference of
assumptions in the hadron phase.
The classes CE and HC reproduce
the pion data well.
On the contrary, results from the second class, PCE, 
deviate from these hydrodynamic results and
experimental data.
To claim the discovery of perfect fluidity,
we need to understand the difference among hydrodynamic
results.
$v_2$ is roughly proportional to $p_T$
in low $p_T$ region for pions.
In such a case, the slope of $v_2(p_T)$
can be approximated by
$v_2/\langle p_T \rangle$.
Integrated $v_2$ is generated in the early
stage of collisions. Whereas \textit{differential}
$v_2$ can be sensitive to the late hadronic stage
since mean transverse momentum $\langle p_T \rangle$ continues to vary.
\begin{figure}[htb]
\begin{minipage}[t]{80mm}
\includegraphics[width=75mm]{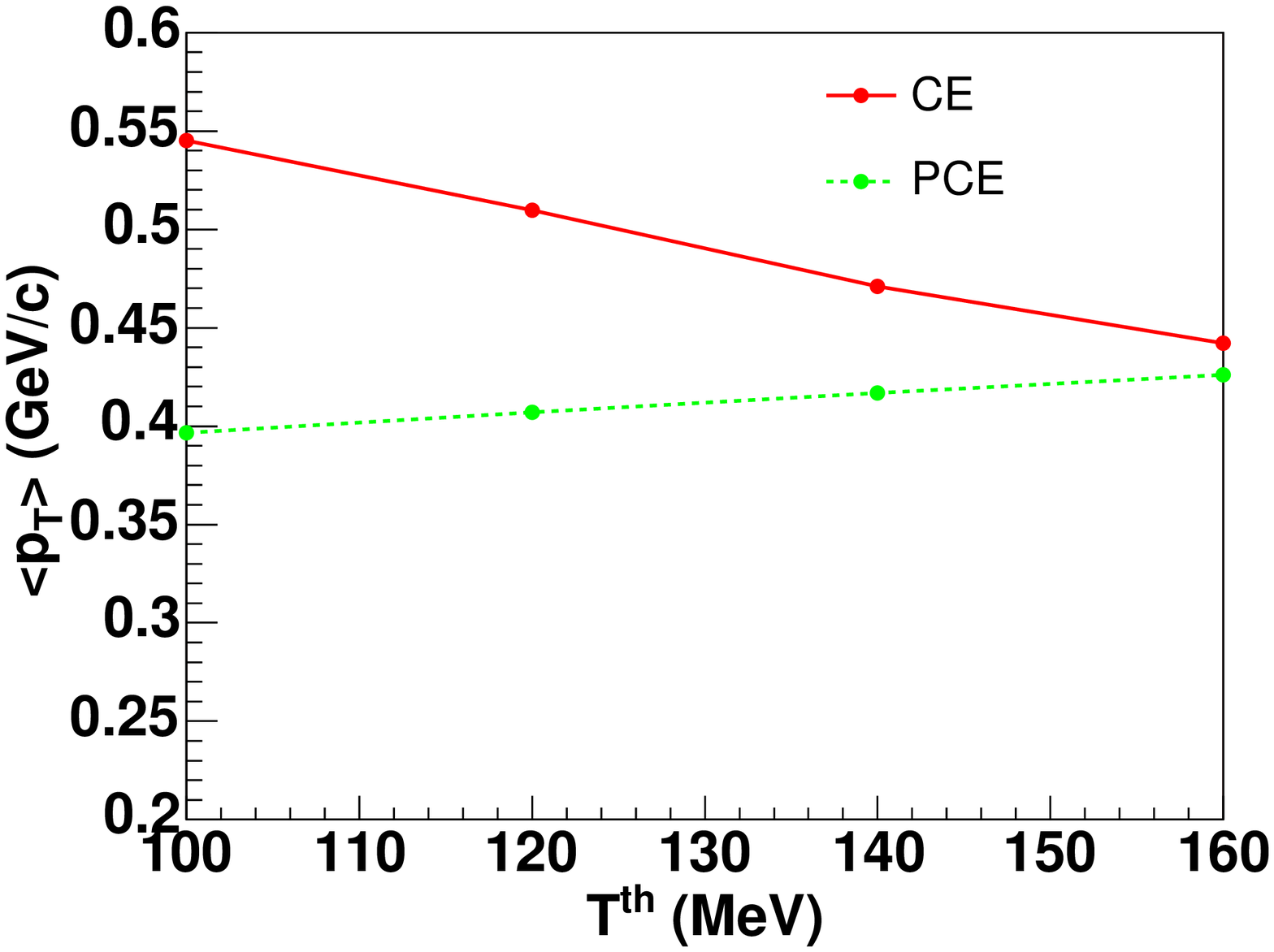}
\end{minipage}
\hspace{\fill}
\begin{minipage}[t]{75mm}
\includegraphics[width=74mm]{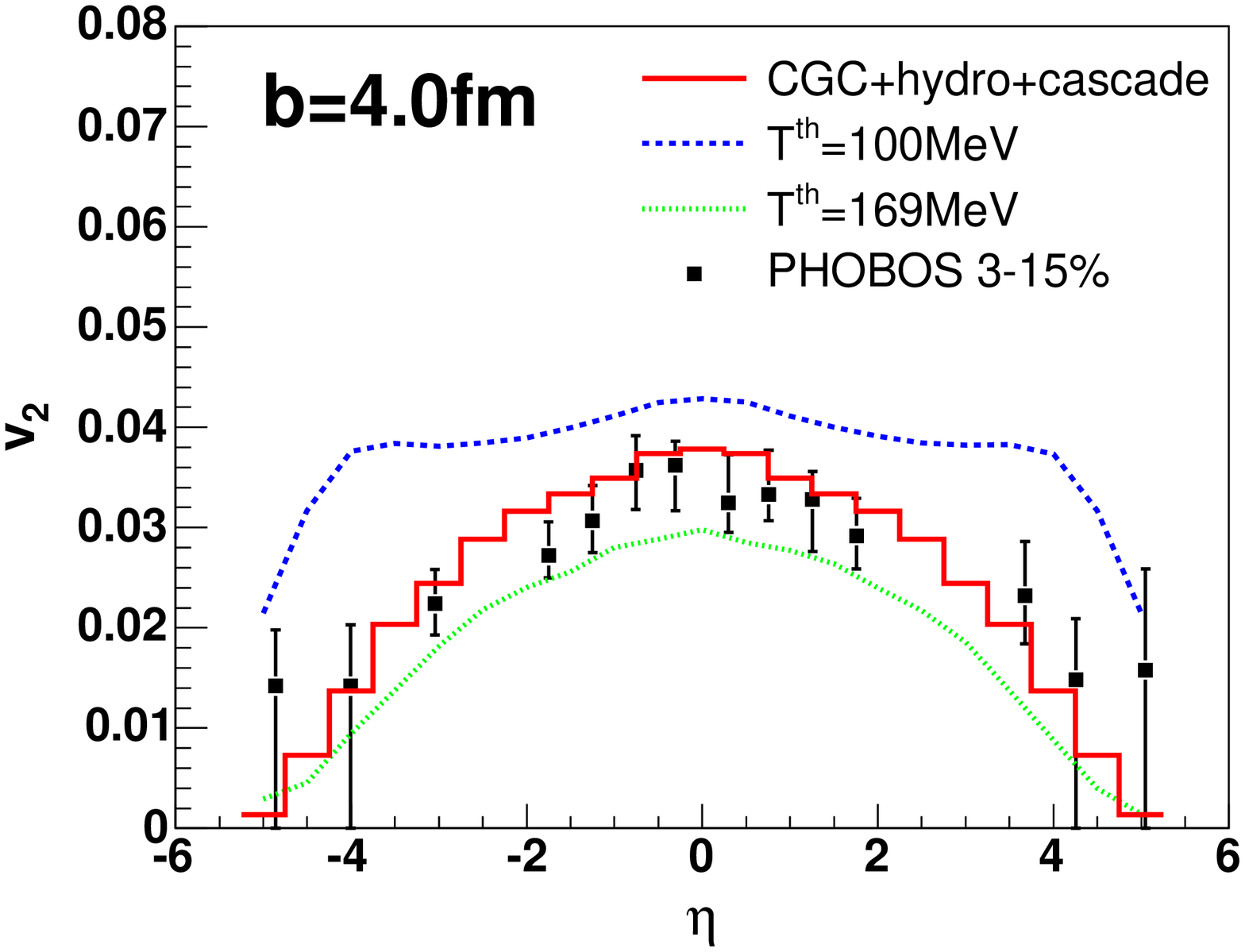}
\caption{(Left) 
Mean transverse momentum for
pions as a function of thermal
freezeout temperature at $b=5$ fm in $\sqrt{s_{NN}}=200$ GeV
Au+Au collisions.
Solid (dashed) line shows a result with
an assumption of an ideal, chemical equilibrium (chemically frozen)
hadronic fluid.
(Right) Pseudorapidity dependence of elliptic flow
from hydro and hydro+hadronic cascade models are
compared with data \cite{Back:2002gz}.
}
\label{fig:v2eta}
\end{minipage}
\end{figure}
In Fig.~\ref{fig:v2eta} (left),
thermal freezeout temperature $T^{\mathrm{th}}$ dependence of
$\langle p_T \rangle$ for pions including
contribution from resonance decays
are calculated at a RHIC energy by utilizing
hydrodynamic simulations \cite{hg}.
$\langle p_T \rangle$ for pions 
in the chemically frozen hadronic fluid
decreases with decreasing $T^{\mathrm{th}}$.
This is due to longitudinal $pdV$ work done by fluid elements.
Whereas $\langle p_T \rangle$ in chemical equilibrium
increases during expansion.
The total number of particles in a fluid element
decreases due to the assumptions of
 chemical equilibrium and entropy conservation in the
hadron phase.
Then the total energy is distributed among the smaller number
of particles as a fluid element expands.
This is the reason why the different behavior 
of $\langle p_T \rangle$ appears
according to the assumption of chemical equilibrium/freezeout.
Under the chemical equilibrium assumption,
increase of $\langle p_T \rangle$ is
commonly utilized to fix  
$T^{\mathrm{th}}$
by fitting $p_T$ slope.
However, this is attained only
by neglecting
particle ratio.
On the other hand, the slope of differential elliptic flow
$v_2/\langle p_T \rangle$
is reproduced by canceling increase behaviors of 
both $v_2$ and $\langle p_T \rangle$ \cite{hg}.
Agreement of 
the CE model
with $p_T$ spectra and $v_2(p_T)$ data
is merely an accident in the sense that this model
makes full use of neglecting particle ratio
to reproduce them.
So the HC model turns out to be the only
model which reproduces \textit{properly}
particle ratio, $p_T$ spectra and $v_2(p_T)$ data.
Therefore a picture of 
the dissipative hadronic corona
together with
the perfect fluid sQGP core 
is consistent with these
experimental data observed at RHIC.

According to the above discussion,
we incorporate a hadronic cascade model, JAM \cite{jam},
into our dynamical framework \cite{hiranonara}.
Details of the hybrid model is found in Ref.~\cite{HHKLN}.
Figure 1 (right) shows pseudorapidity dependences of $v_2$
from this hybrid model 
and ideal 3D hydrodynamics with $T^{\mathrm{th}}=100$ and 169 MeV.
Here critical temperature
and chemical freezeout temperature
are taken as being 
$T_c = T^{\mathrm{ch}}=170$ MeV
in the hydrodynamic model.
In the hybrid model,
the switching temperature
from a hydrodynamic description to a kinetic one
is taken as $T_{\mathrm{sw}}=169$ MeV.
Ideal hydrodynamics with $T^{\mathrm{th}}=100$ MeV which is so chosen
to generate enough radial flow gives
a trapezoidal shape of $v_2(\eta)$ \cite{hirano3d}.
A large deviation between data \cite{Back:2002gz}
and the ideal hydrodynamic result
is seen especially in forward/backward rapidity regions.
When hadronic rescattering effects are taken through
the hadronic cascade model
instead of perfect fluid description of the hadron phase,
$v_2$ is not so generated in the forward region due to the
dissipation and, eventually, is consistent with
the data. 
So the perfect fluid sQGP core and the dissipative hadronic
corona picture works well also in the forward region.
It should be noted that, recently, this conclusion is
found to depend on centralities and how to initialize hydrodynamic
fields: There is a room for
initial QGP viscous effects in peripheral collisions
in CGC initial conditions. For a detailed discussion, see Ref.~\cite{HHKLN}.

We can establish a new picture of space-time evolution
of produced matter from 
a careful comparison of hydrodynamic
results with experimental data
observed at RHIC.
What is the physics behind this picture?
$\eta/s$ is known to be a good measure to see
the effect of viscosity where $\eta$
is the shear viscosity and $s$ is the entropy density.
$\eta/s$ might be small in the QGP phase,
which is comparable 
with the minimum value $1/4\pi$ \cite{Son}, 
and the perfect fluid assumption
can be valid. While $\eta/s$ becomes huge
in the hadron phase and the dissipation
cannot be neglected.
Shear viscosities of both phases 
are found to give $\eta \sim 0.1$ GeV/fm$^2$
around $T_c$ \cite{hg}.
So shear viscosity itself appears to increase with temperature
monotonically.
The ``perfect fluid'' property of the sQGP is thus not
due to a sudden reduction of the viscosity
at the critical temperature $T_c$, but to
a sudden increase of the entropy density 
of QCD matter and is therefore an
important signature of deconfinement.




\begin{theacknowledgments}
This work was supported in part by the United States
Department of Energy under
Grant No.~DE-FG02-93ER40764.
The author would like to thank U.~Heinz, D.~Kharzeev,
M.~Gyulassy, R.~Lacey and Y.~Nara
for collaboration and fruitful discussion.
\end{theacknowledgments}



\bibliographystyle{aipproc}   


\begin{thebibliography}{9}


 \bibitem{Huovinen:2003fa}
  P.~Huovinen,
in \textit{Quark Gluon Plasma 3},
  (World Scientific Pub., 2004, eds. R.C. Hwa, X.N. Wang) p.~600;
  P.~F.~Kolb and U.~Heinz,
\textit{ibid.} p.~634;
  D.~Teaney, J.~Lauret and E.~V.~Shuryak,
  nucl-th/0110037;
  T.~Hirano,
  Acta Phys.\ Polon.\ B {\bf 36}, 187 (2005);
Y.~Hama {\it et al.}, hep-ph/0510096.


\bibitem{Ackermann:2000tr}
  K.~H.~Ackermann {\it et al.}  [STAR Collaboration],
  Phys.\ Rev.\ Lett.\  {\bf 86}, 402 (2001);
  C.~Adler {\it et al.}  [STAR Collaboration],
\textit{ibid.} {\bf 87}, 182301 (2001);
{\bf 89}, 132301 (2002);
  Phys.\ Rev.\ C {\bf 66}, 034904 (2002);
  J.~Adams {\it et al.}  [STAR Collaboration],
  nucl-ex/0409033.

\bibitem{Adcox:2002ms}
  K.~Adcox {\it et al.}  [PHENIX Collaboration],
  Phys.\ Rev.\ Lett.\  {\bf 89}, 212301 (2002);
  S.~S.~Adler {\it et al.}  [PHENIX Collaboration],
\textit{ibid.}
  {\bf 91}, 182301 (2003);

\bibitem{Back:2002gz}
  B.~B.~Back {\it et al.}  [PHOBOS Collaboration],
  Phys.\ Rev.\ Lett.\  {\bf 89}, 222301 (2002);
  nucl-ex/0406021;
  nucl-ex/0407012.

\bibitem{Ito} E.~Johnson {\it et al.}
[BRAHMS Collaboration], these proceedings.

\bibitem{BNL} \url{http://www.bnl.gov/bnlweb/pubaf/pr/PR\_display.asp?prID=05-38}




\bibitem{Adcox:2004mh}
  K.~Adcox {\it et al.}  [PHENIX Collaboration],
  nucl-ex/0410003.

\bibitem{hg} T.~Hirano and M.~Gyulassy, nucl-th/0506049.


\bibitem{jam} Y.~Nara \textit{et al.},
 Phys. Rev. C {\bf 61}, 024901 (2000).

\bibitem{hiranonara}
  T.~Hirano and Y.~Nara,
  Nucl.\ Phys.\ A {\bf 743}, 305 (2004).

\bibitem{HHKLN}
T.~Hirano \textit{et al.}, nucl-th/0511046.

\bibitem{hirano3d}
  T.~Hirano,
  Phys.\ Rev.\ C {\bf 65}, 011901 (2002);
 T.~Hirano and K.~Tsuda,
\textit{ibid.} {\bf 66}, 054905 (2002).

\bibitem{Son}
  P.~Kovtun, D.~T.~Son and A.~O.~Starinets,
  Phys.\ Rev.\ Lett.\  {\bf 94}, 111601 (2005);
  G.~Policastro, D.~T.~Son and A.~O.~Starinets,
\textit{ibid.}  
 {\bf 87}, 081601 (2001).


%
%
%

\end{thebibliography}



\end{document}